\documentclass[twoside,12pt]{article}
\usepackage{graphicx}
\usepackage{amsmath}
\usepackage{amssymb,dsfont}
\begin{document}
\title{ \vspace{1cm} A lattice study of the strangeness content
of the nucleon}
\author{G. S.\ Bali,$^1$, S.\ Collins,$^1$ M.\ G\"ockeler,$^1$
R.\ Horsley,$^2$ Y.\ Nakamura,$^3$\\
A.\ Nobile,$^1$ D.\ Pleiter,$^{4,1}$
P. E. L.\ Rakow,$^5$ A.\ Sternbeck,$^1$\\
A.\ Sch\"afer,$^1$
G.\ Schierholz,$^6$ J. M.\ Zanotti$^7$\\
(QCDSF Collaboration)\\
\\
${}^1$Institut f\"ur Theoretische Physik, Universit\"at Regensburg,\\
93040 Regensburg, Germany\\
${}^2$School of Physics, University of Edinburgh, Edinburgh EH9 3JZ, UK\\
${}^3$RIKEN Advanced Institute for Computational Science, Kobe,\\
Hyogo 650-0047, Japan\\
${}^4$JSC, Research Center J\"{u}lich, 52425 J\"{u}lich, Germany\\
${}^5$Theoretical Physics Division, Department of Mathematical Sciences,\\
University of Liverpool, Liverpool L69 3BX, UK\\
${}^6$Deutsches Elektronen-Synchrotron DESY, 22603 Hamburg, Germany\\
${}^7$CSSM,
School of Chemistry \& Physics, University of Adelaide,\\
South Australia 5005, Australia}
\maketitle
\begin{abstract}
We determine the quark contributions to the nucleon spin
$\Delta s$, $\Delta u$ and $\Delta d$ as well
as their contributions to
the nucleon mass, the $\sigma$-terms.
This is done by computing both,
the quark line connected and
disconnected contributions to the respective matrix
elements, using the non-perturbatively improved
Sheikholeslami-Wohlert Wilson Fermionic action.
We simulate $n_{\mathrm{F}}=2$ mass degenerate sea quarks 
with a pion mass of about 285~MeV and a lattice
spacing $a\approx 0.073$~fm.
The renormalization of the matrix elements
involves mixing between contributions from different quark flavours.
The pion-nucleon $\sigma$-term is extrapolated to
physical quark masses exploiting the sea quark mass dependence
of the nucleon mass.
We obtain the renormalized value $\sigma_{\pi\mathrm{N}}=(38\pm 12)$ MeV
at the physical point
and the strangeness fraction $f_{T_s}=\sigma_s/m_{\mathrm{N}}=
0.012(14)^{+10}_{-3}$ at our larger
than physical sea quark mass.
For the strangeness contribution to the
nucleon spin
we obtain $\Delta s^{\overline{MS}}(\sqrt{7.4}\,\mathrm{GeV})=-0.020(10)(2)$.
\end{abstract}
\section{Introduction}
Most of the nucleon's mass is 
generated by the spontaneous breaking of chiral symmetry
and only a small part can be attributed
directly to the masses of its valence and sea quarks.
The quantities
\begin{equation}
\label{eq:ftq}
f_{T_q}=m_q\langle N|\bar{q}{q}|N\rangle/m_{\mathrm{N}}=\sigma_q/m_N
\end{equation}
parameterize the fractions
of the nucleon mass $m_{\mathrm{N}}$ that are carried by
quarks of flavour $q$.
These scalar matrix elements
also determine the coupling strength of the
Standard Model Higgs boson (or of any similar scalar particle)
at zero recoil to the nucleon.
This then might couple to heavy particles, some of
which are dark matter candidates~\cite{Ellis:2009ai}.
The combination $m_{\mathrm{N}}\sum_qf_{T_q}$, $q\in\{u,d,s\}$,
appears quadratically in the cross section
that is proportional to $|f_{\mathrm{N}}|^2$, where
\begin{equation}
\label{eq:fn}
f_{\mathrm{N}}=m_{\mathrm{N}}\left(\sum_{q\in\{u,d,s\}}\!\!\!\!\!\!f_{T_q}\frac{\alpha_q}{m_q}
+\frac{2}{9n_h}f_{T_G}\!\!\!\!\!\!\sum_{q\in \{c,b,t,\ldots\}}\!\!\frac{\alpha_q}{m_q}\right)\,,\end{equation}
with the couplings
$\alpha_q\propto m_q/m_W$. $n_h$ denotes the number
of heavy quark flavours. Note that due to the trace anomaly
of the energy momentum tensor one obtains
\begin{equation}
\label{eq:ftg}
f_{T_G}= 1-\!\!\!\!\sum_{q\in\{u,d,s\}}\!\!\!\!f_{T_q}\,,
\end{equation}
so that the coupling $f_{\mathrm{N}}$ only depends
mildly on the
number and properties
of (discovered and not yet discovered) heavy quark
flavours~\cite{Shifman:1978zn}.

The light quark contribution, the pion-nucleon $\sigma$-term, is defined as
\begin{equation}
\label{eq:pin}
\sigma_{\pi\mathrm{N}}=\sigma_u+\sigma_d=
m_u\frac{\partial m_{\mathrm{N}}}{\partial m_u}
+m_d\frac{\partial m_{\mathrm{N}}}{\partial m_d}\approx
\left.m^2_{\mathrm{PS}}\frac{dm_{\mathrm{N}}}{dm_{\mathrm{PS}}^2}\right|_{m_{\mathrm{PS}}=m_{\pi}}
\!\!\!\!\!.
\end{equation}
From dispersive analyses of pion-nucleon scattering data,
the values~\cite{Gasser:1990ce} $\sigma_{\pi N}=45(8)$~MeV
and~\cite{Pavan:2001wz} $\sigma_{\pi\mathrm{N}}=64(7)$~MeV were obtained
while a recent covariant baryon chiral perturbation theory
(B$\chi$PT) analysis of the available
scattering and pionic atom data~\cite{Alarcon:2011zs} resulted in the
estimate
$\sigma_{\pi\mathrm{N}}=59(7)$~MeV.

Not only do the quarks contribute a tiny fraction
to the nucleon's mass, even the nucleon spin is mostly
carried by the gluons. This spin can be factorized
into a quark spin contribution
$\Delta\Sigma$, a quark angular
momentum contribution $L_q$ and gluonic contributions
$\Delta G$ and $L_G$ (for spin and angular momentum):
\begin{equation}
\frac12=\frac12\Delta\Sigma+L_q+\Delta G+L_G\,.
\end{equation}
In the na\"{\i}ve non-relativistic SU(6) quark model, $\Delta \Sigma = 1$,
with vanishing angular momentum and gluon contributions.
In this case there will be no strangeness contribution
$\Delta s$ to
$\Delta \Sigma = \Delta u+\Delta d+\Delta s +\cdots$
where, in our notation, $\Delta q$  contains both, the spin of the quarks
$q$ and of the antiquarks $\bar{q}$. Experimentally,
$\Delta s$ is obtained by integrating the strangeness contribution
to the spin structure function $g_1$ over the momentum fraction $x$.
The integral over the range in which data exists usually
agrees with zero, see e.g. new COMPASS
data~\cite{arXiv:1001.4654} for $x\geq 0.004$,
while global analyses tend to obtain values
$\Delta s\approx -0.12$~\cite{arXiv:1010.0574,arXiv:0904.3821}.

Here, we directly compute the quark line disconnected
(and connected) contributions to the scalar and axial matrix
elements that appear in the above
quantities~\cite{arXiv:1111.1600,arXiv:1011.2194}.
Other recent direct lattice determinations of these quantities
include Refs.~\cite{arXiv:1111.5426,arXiv:1108.2473,arXiv:1012.0562,arXiv:1011.6058,arXiv:1011.1964,arXiv:0903.3232}.
We will first describe the methods used, before we present
results on the quark spin contributions and the scalar matrix elements.

\section{Methods and simulation parameters}
We simulate $n_{\mathrm{F}}=2$ non-perturbatively
improved Sheikholeslami-Wohlert Fermions, using the
Wilson gauge action, at $\beta=5.29$
and $\kappa=\kappa_{ud}=0.13632$. 
Setting the scale from the chirally extrapolated nucleon mass~\cite{sternbeck},
we obtain the lattice spacing
$a^{-1}=2.71(2)(7)\,\mathrm{GeV}$,
where the errors are statistical and from the extrapolation,
respectively.

\begin{figure}[t]
\centerline{\includegraphics[height=.6\textwidth,angle=270,clip=]{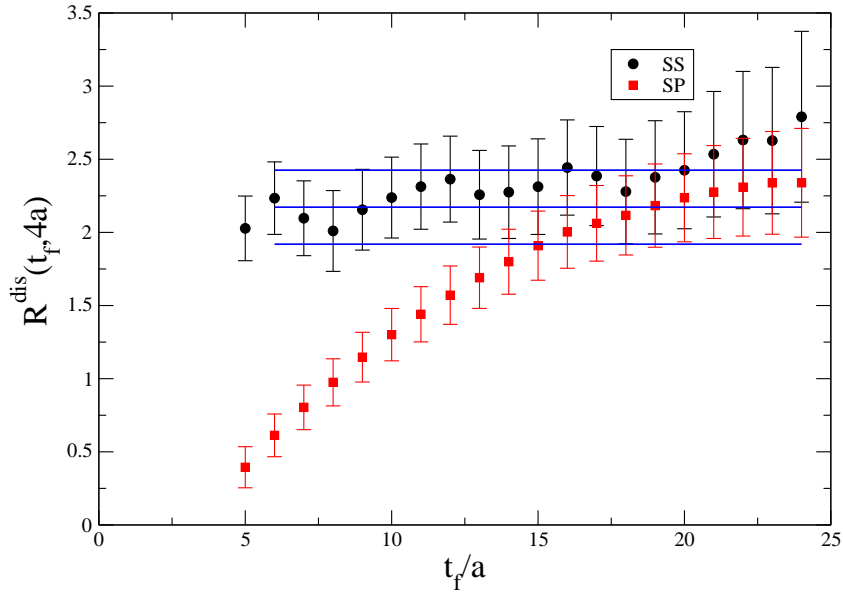}}
\caption{Dependence of $R^{\mathrm{dis}}$ on $t_{\mathrm{f}}$ for
the scalar strangeness matrix element on the $40^364$ volume
for smeared-smeared (SS) and smeared-point (SP) two-point functions,
together with the fit result.
\label{fig:ratio2}}
\end{figure}

We realize two additional
valence $\kappa$ values,
$\kappa_m=0.13609$ and $\kappa_s=0.13550$. The corresponding
pseudoscalar masses are
$m_{\mathrm{PS},ud}=285(3)(7)\,\mathrm{MeV}$,
$m_{\mathrm{PS},m}=449(3)(11)\,\mathrm{MeV}$ and
$m_{\mathrm{PS},s}=720(5)(18)\,\mathrm{MeV}$.
The strange quark mass was fixed so that the $m_{\mathrm{PS},s}$ value
is close to the mass of
a hypothetical strange-antistrange pseudoscalar meson:
$(m_{K^{\pm}}^2+m_{K^0}^2-m_{\pi^{\pm}}^2)^{1/2}\approx
686.9$~MeV.
We investigate volumes of
$32^364$ and $40^364$ lattice points,
i.e., $Lm_{\mathrm{PS}}= 3.36$ and 4.20,
respectively, where the largest spatial lattice extent is
$L\approx 2.91$~fm.

The matrix elements of interest are extracted from the large
time behaviour of ratios of three-point functions over
two-point functions where we create a proton at a time $t_0=0$
and destroy it at zero momentum at a time $t_{\mathrm{f}}$.
At an intermediate time $0<t<t_{\mathrm{f}}$ the current of interest
is inserted. The three-point function contains both, a
quark line connected and a quark line disconnected contribution.
For the example of the axial current the ratios to be calculated read
\begin{equation}
R^{\mathrm{con}}(t_{\mathrm{f}},t) =  \frac{\langle\Gamma_{\rm pol}^{\alpha\beta} C^{\beta\alpha}_{3pt}(t_{\mathrm{f}},t) \rangle}{\langle \Gamma_{\rm unpol}^{\alpha\beta}C^{\beta\alpha}_{2pt}(t_{\mathrm{f}}) \rangle} \,,\quad
R^{\mathrm{dis}}(t_{\mathrm{f}},t) =  -\frac{\langle \Gamma_{\rm pol}^{\alpha\beta}C^{\beta\alpha}_{2pt}(t_{\mathrm{f}}) \sum_{\mathbf{x}}\mathrm{Tr}(\gamma_j\gamma_5 M^{-1}(\mathbf{x},t;\mathbf{x},t))\rangle}{\langle \Gamma_{\rm unpol}^{\alpha\beta} C^{\beta\alpha}_{2pt}(t_{\mathrm{f}})\rangle}\,,
\end{equation}
where $M$ is the lattice Dirac operator,
$\Gamma_{\mathrm{unpol}}=(\mathds{1}+\gamma_4)/2$ a parity projector
and $\Gamma_{\mathrm{pol}}=i\gamma_j\gamma_5\Gamma_{\mathrm{unpol}}$
projects out the difference between the two polarizations
(in direction $\hat{\boldsymbol{\jmath}})$. We average over
$j=1,2,3$ to increase statistics. For the scalar case
we have to replace $\gamma_j\gamma_5\mapsto \mathds{1}$,
$\Gamma_{\mathrm{pol}}\mapsto\Gamma_{\mathrm{unpol}}$ and add
the vacuum condensate
$\langle\sum_{\mathbf{x}}\mathrm{Tr}\,M^{-1}(\mathbf{x},t;\mathbf{x},t)\rangle$
to $R^{\mathrm{dis}}$ above.
For the up and down quark matrix elements we
compute the sum of connected and disconnected terms
while only $R^{\mathrm{dis}}$ contributes to $\Delta s$ and $\sigma_s$.

The disconnected contribution is computed with the methods described
in~\cite{arXiv:0910.3970,arXiv:1111.1600,arXiv:1011.2194} where
we fix $t=4a\approx 0.29$~fm and vary $t_{\mathrm{f}}$.
Employing optimized sink and source smearing
we find the asymptotic limit
to be effectively reached for $t_{\mathrm{f}}\geq 5a$
and fit the ratios to a constant for
$t_{\mathrm{f}}\geq 6a\approx 0.44$~fm, see Fig.~\ref{fig:ratio2} for an example.

\begin{figure}[t]
\centerline{
\includegraphics[width=.6\textwidth,clip=]{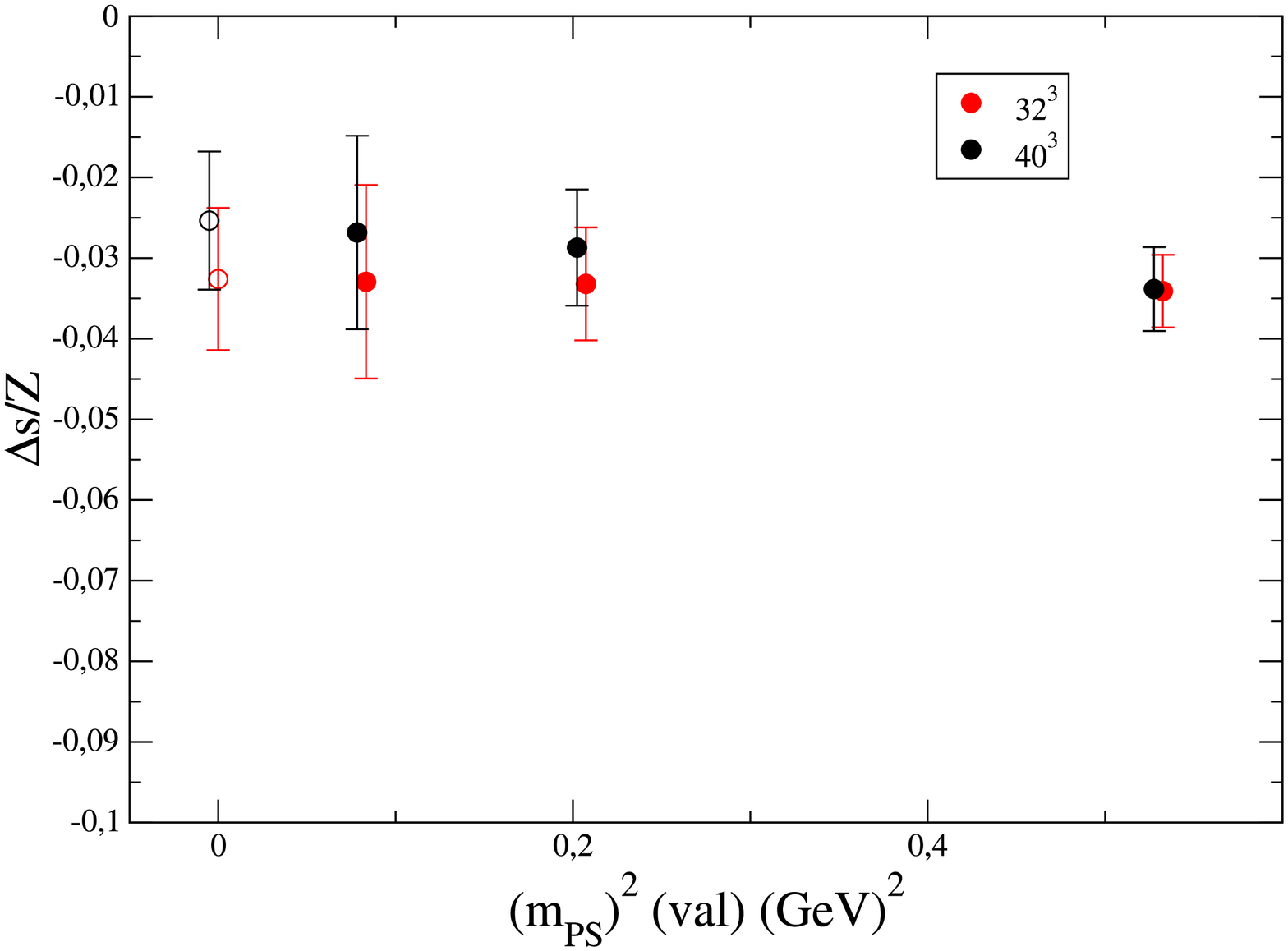}}
\caption{Volume and (light) valence quark mass dependence
of the unrenormalized $\Delta s^{\mathrm{lat}}$.\label{fig:deltas}}
\end{figure}

\begin{table}
\begin{center}
\begin{minipage}[t]{16.5 cm}
\caption{The connected and disconnected 
contributions to $\Delta q^{\mathrm{lat}}$ of the proton
on the $40^364$ volume ($L\approx 2.91$~fm)
as well as the renormalized spin content at a scale
$\mu\approx \sqrt{7.4}$~GeV.
\label{tab:spin}}
\end{minipage}
\begin{tabular}{c|c|c|c}
\hline
$q$&$\Delta q^{\mathrm{lat}}_{\rm con}$&$\Delta q^{\mathrm{lat}}_{\rm dis}$&
$\Delta q^{\mathrm{\overline{MS}}}(\mu)$\\\hline
$u$&~1.071(15)&-0.049(17)&0.787(18)(2)\\
$d$&-0.369( 9)&-0.049(17)&-0.319(15)(2)\\
$s$& 0        &-0.027(12)&-0.020(10)(2)\\\hline
\end{tabular}
\end{center}
\end{table}

\section{The quark contributions to the proton spin}
Non-singlet axial currents renormalize with a
renormalization factor $Z_A^{ns}(a)$ that only depends
on the lattice spacing. This was determined
non-perturbatively for the action and lattice spacing
in use~\cite{Gockeler:2010yr}: $Z_A^{ns}=0.76485(64)(73)$.
However, due to the axial anomaly, the renormalization
constant of singlet currents, $Z_A^s(\mu,a)$, acquires an
anomalous dimension~\cite{Kodaira:1979pa} and will
depend on the scheme and scale used in the continuum.
This has been determined perturbatively~\cite{Skouroupathis:2008mf}
and the result for the
conversion to the $\overline{MS}$ scheme reads
\begin{equation}
z(\mu,a):=Z_A^s(\mu,a)-Z_A^{ns}(a)=C_{\mathrm{F}}n_{\mathrm{F}}\left[15.8380(8)
-6\ln(a^2\mu^2)\right]\left(\frac{\alpha_s}{4\pi}\right)^2+
\mathcal{O}(\alpha_s^3)\,.
\end{equation}
We extract an improved coupling from the measured plaquette, set
$\mu=a^{-1}$
and allow for a 50~\% systematic error on 
$z(\sqrt{7.4}\,\mbox{GeV})=0.0055(1)(27)$.
As can be seen from the small anomalous dimension, the
scale dependence of $z(\mu)$ is quite mild.
Perturbative $\mathcal{O}(a)$ improvement is implemented~\cite{Capitani:2000xi}
to $\mathcal{O}(\alpha)$, where again we allow for a 50~\% systematic error.

In the $n_{\mathrm{F}}=1+1+1$ theory the matrix elements
renormalize as follows.
\begin{align}\label{eq:dq1}
\Delta\Sigma^{\overline{MS}}(\mu)=(\Delta u +\Delta d +\Delta s)^{\overline{MS}}(\mu)
&=Z^s_A(\mu,a)(\Delta u +\Delta d +\Delta s)^{\mathrm{lat}}(a)\,,\\\label{eq:dq2}
a_8=\Delta T_8=(\Delta u +\Delta d -2\Delta s)^{\overline{MS}}
&=Z^{ns}_A(a)(\Delta u +\Delta d -2\Delta s)^{\mathrm{lat}}(a)\,,\\\label{eq:dq3}
g_A=\Delta T_3=(\Delta u -\Delta d )^{\overline{MS}}
&=Z^{ns}_A(a)(\Delta u -\Delta d)^{\mathrm{lat}}(a)\,.
\end{align}
We employ $n_{\mathrm{F}}=2$ mass-degenerate sea quarks
so that our singlet current is $\Delta u +\Delta d$ instead.
This modifies the renormalization
pattern~\cite{arXiv:1011.2194}
\begin{equation}\label{eq:dqren}
\left(
\begin{array}{c}
\Delta u(\mu)\\
\Delta d(\mu)\\
\Delta s(\mu)\end{array}\right)^{\overline{MS}}
=
\left(
\begin{array}{ccc}
Z_A^{ns}(a)+\frac{z(\mu,a)}{2}&\frac{z(\mu,a)}{2}&0\\
\frac{z(\mu,a)}{2}&Z_A^{ns}(a)+\frac{z(\mu,a)}{2}&0\\
\frac{z(\mu,a)}{2}&\frac{z(\mu,a)}{2}&Z_A^{ns}(a)\end{array}\right)
\left(
\begin{array}{c}
\Delta u(a)\\
\Delta d(a)\\
\Delta s(a)\end{array}\right)^{\mathrm{lat}}\,,
\end{equation}
where $\Delta s^{\overline{MS}}$ receives light quark
contributions
but the $\Delta u^{\overline{MS}}$ and $\Delta d^{\overline{MS}}$ remain unaffected by the
(quenched) strange quark. We remark that unitarity is violated,
due to the partial quenching.

In Fig.~\ref{fig:deltas} we display the volume and
(light) valence quark mass dependence of our
unrenormalized $\Delta s^{\mathrm{lat}}$.
There are no significant finite size effects and we take
the independence on the valence quark mass as an
indication that our result may also approximately apply to
physical light quark masses. 
We display our $\mathcal{O}(a)$ improved
results with statistical and systematic errors
in Table~\ref{tab:spin}.
The $\Delta u^{\overline{MS}}$ and $\Delta d^{\overline{MS}}$
values are
reduced by about 0.035, due to the sea
quark contributions while $\Delta s^{\overline{MS}}$ increases
by 0.002 ($<10$~\%), due to the mixing with light quark
flavours.

Note that we find $g_A\approx 1.11(2)$.
This underestimation of the  value
$g_A=1.267(4)$ from
neutron $\beta$-decays
can probably be explained by our twice as heavy as
physical pseudoscalar meson.
Our main finding is a small negative
$\Delta s^{\overline{MS}}(\sqrt{7.4}\,\mathrm{GeV})=-0.020(11)$
that is unlikely to decrease significantly if the
sea quark mass is reduced: the mixing effects on the
renormalization are small, in spite of the comparatively
large $\Delta u$ and $\Delta d$ values, and so
is the dependence on the light valence
quark mass, see Fig.~\ref{fig:deltas}.

\section{The light and strange $\sigma$-terms}
In massless schemes that preserve the chiral
symmetry the combinations $m_q\bar{q}q$
are invariant under scale transformations
(up to lattice artefacts). This 
operator identity holds for expectation values, independent
of the external state. The Wilson action
however explicitly breaks the chiral symmetry so that
singlet and non-singlet flavour combinations will
renormalize differently. It turns out that at our lattice spacing
the ratio between the two renormalization
factors deviates from unity by as much
as 40~\%. Consequently,
the renormalized strangeness matrix element receives
large subtractions from light quark contributions.
For details, see~\cite{arXiv:1111.1600} and references therein.
Moreover, without
considering mixing with the gluonic operator
$a GG$ (that we have not determined),
we are not able to implement $\mathcal{O}(a)$
improvement. Note that due to the trace anomaly
of the energy-momentum tensor such gluonic operators
also become relevant for heavy quark masses
in the continuum theory~\cite{Shifman:1978zn}.
We plan to take this effect into account in the future.

\begin{figure}[t]
\centerline{
\includegraphics[width=.65\textwidth,clip=]{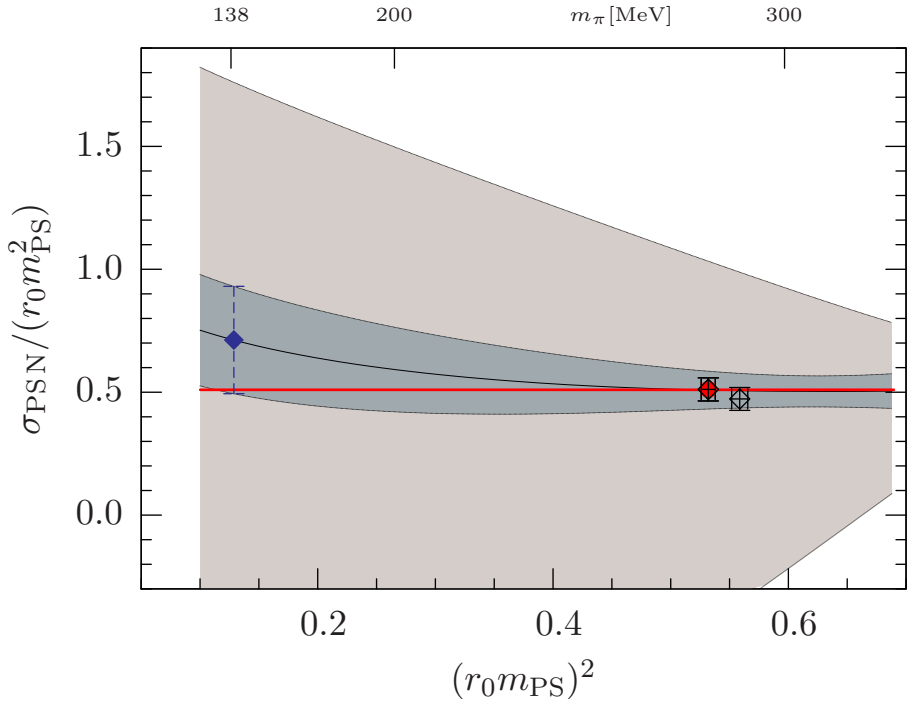}}
\caption{Extrapolation of $\sigma_{\mathrm{PS\,N}}/m_{\mathrm{PS}}^2$
to the physical point~\protect\cite{sternbeck} using covariant B$\chi$PT
for the $40^364$ volume (solid symbol).
The broad error band is obtained from nucleon mass data alone.
The horizontal line is the leading order expectation and the
open symbol our result for the $32^364$ volume.\label{fig:sigma}}
\end{figure}
Again we find no significant finite size effects and
obtain
$\sigma_{\mathrm{PS\,N}}(
m_{\mathrm{PS}}\approx 285\,\mbox{MeV})=106(11)(3)$~MeV. Recently,
the ETM Collaboration~\cite{arXiv:1111.5426} reported the value
$\sigma_{\mathrm{PS\,N}}(
m_{\mathrm{PS}}\approx 380\,\mbox{MeV})
=150(1)(10)$~MeV, using a
different lattice action.
If we assume the leading order chiral behaviour
$dm_{\mathrm{N}}/dm_{\mathrm{PS}}^2=\mathrm{const}$ then Eq.~(\ref{eq:pin})
suggests to rescale the results according to
the ratio of the squared pseudoscalar masses $\approx 1.78$.
However, $1.42(19)<1.78$: higher order chiral corrections are relevant.
Using additional nucleon mass data we extrapolate our value to
the physical point~\cite{sternbeck} and obtain~\cite{arXiv:1111.1600}
\begin{equation}
\sigma^{\mathrm{phys}}_{\pi\mathrm{N}}=(38\pm 12)\,\mathrm{MeV}\,,
\end{equation}
where the dominant error is from the chiral extrapolation,
see Fig.~\ref{fig:sigma}.

We are also able to compute the strangeness and gluon contributions
to the nucleon mass, see Eqs.~(\ref{eq:ftq}) and (\ref{eq:ftg}):
\begin{equation}
f_{T_s}=0.012(14)^{+10}_{-3}\,,\quad
f_{T_G}=0.951^{+20}_{-27}\,.
\end{equation}
The light and strange quarks
contribute a fraction between 3~\% and 8~\% to the
nucleon mass. The large uncertainty on $f_{T_s}$
or, equivalently, $\sigma_s=12^{+23}_{-16}$~MeV is due to large cancellations
in the renormalization. To reduce these, with Wilson Fermions,
one will need to simulate at finer lattice spacings where
differences between singlet and non-singlet renormalization
constants become smaller. We are able to state a 95~\% confidence
level upper limit on the phenomenologically relevant $y$-ratio:
$y<0.14$.

\section{Conclusion}
We calculated disconnected contributions to
the proton structure. We find a pion-nucleon
$\sigma$-term $\sigma_{\pi N}=38(12)$ MeV at the
physical point and an upper limit on the
ratio of scalar strangeness over light quark content
of $y<0.14$. The renormalized
light sea quark contributions amount to less than
10~\% of the total matrix elements,
both for the spin content and for the
$\sigma$-term. However,
at our lattice spacing, prior to the renormalization, these account for 30~\% of the
bare scalar lattice matrix elements~\cite{arXiv:1111.1600} and need to be taken into account
for Wilson Fermions.

Our small negative value $\Delta s^{\overline{MS}}(\sqrt{7.4}\,\mbox{GeV})=-0.020(11)$ indicates
stronger than na\"{\i}vely expected
violations of SU$(3)_F$ symmetry in weak decays.
This impacts on determinations of polarized
parton distribution functions~~\cite{arXiv:1002.4407,arXiv:1010.0574,arXiv:0904.3821}
where the
constraint on the integrals by the (assumed) proton tensor charge
value $a_8=3F-D$ should probably be relaxed.
A small value of $\Delta s$ was also
reported in~\cite{arXiv:1012.0562},
albeit without renormalization.
In view of these findings, lattice studies of the spin content
of hyperons and of their weak transition matrix elements
seem particularly interesting.
\section*{Acknowledgements}
This work was supported by the European Union
(grant 238353, ITN STRONGnet)
and by the DFG SFB/Transregio 55. S.C.\
is supported by the Claussen-Simon-Foundation (Stifterverband
f\"ur die Deutsche Wissenschaft),
A.St.\ by the EU IRG grant 256594 and
J.Z.\ by the Australian Research Council
grant FT100100005. Computations were performed on the
SFB/TR55 QPACE supercomputers,
the BlueGene/P (JuGene) and the Nehalem cluster (JuRoPA) of the
JSC (J\"ulich), the
IBM BlueGene/L at the EPCC (Edinburgh),
the SGI Altix ICE machines at HLRN (Berlin/Hannover)
and Regensburg's Athene HPC cluster.
The Chroma software suite~\cite{Edwards:2004sx} was used extensively
in this work.

\end{document}